\documentclass[twocolumn,aps,prb,superscriptaddress,amsmath,amssymb]{revtex4-2}
\usepackage{amsmath}
\usepackage{newtxmath}
\usepackage[utf8]{inputenc}
\usepackage{color}
\usepackage{mathrsfs}
\usepackage{bm}
\usepackage{graphicx}
\usepackage[bookmarks=false,
 breaklinks=true,pdfborder={0 0 1},backref=section,colorlinks=true]
 {hyperref}
\hypersetup{
 citecolor=blue,linkcolor=blue,linktocpage=true,urlcolor=blue,pagebackref=false}

\makeatletter
\usepackage{newtxtext}
\usepackage{bm}
\usepackage{siunitx}
\usepackage{tabularx}
\usepackage{appendix}

\usepackage{placeins}

\usepackage{multirow}

\DeclareGraphicsExtensions{.pdf,.png}

\newcommand{\void}[1]{}

\newcommand{\sect}[1]{\par\vspace{3ex}\textit{#1}.---\ignorespaces}

\makeatother

\begin{document}
\title{Interplay between Hund’s rule and Kondo effect in a quantum dot}
\author{Olfa Dani}
\affiliation{Institut für Festkörperphysik, Leibniz Universität Hannover, D-30167
Hanover, Germany}
\author{Johannes C. Bayer}
\affiliation{Institut für Festkörperphysik, Leibniz Universität Hannover, D-30167
Hanover, Germany}
\affiliation{Physikalisch-Technische Bundesanstalt, 38116 Braunschweig, Germany}
\author{Timo Wagner}
\affiliation{Institut für Festkörperphysik, Leibniz Universität Hannover, D-30167
Hanover, Germany}
\author{Gertrud Zwicknagl}
\affiliation{Institut für Mathematische Physik, Technische Universität Braunschweig,
D-38106 Braunschweig, Germany}
\author{Rolf J. Haug}
\affiliation{Institut für Festkörperphysik, Leibniz Universität Hannover, D-30167
Hanover, Germany}
\date{\today}
\begin{abstract}
The interaction between localized spins on a quantum dot and free
electrons in the reservoirs forms a many-particle entangled system
giving rise to the Kondo effect. Here, we investigate electron transport
in the third shell of a gate-defined GaAs quantum dot. The addition
energy shows a maximum at half-filling of the shell which can be described
analytically with Hund’s rule exchange interaction. For 7 to 11 electrons
occupying the quantum dot Zero-bias anomalies characteristic for the
Kondo effect are observed, but with unexpected widths. Here, the quantum
dot has to be described as a multi-orbital Kondo impurity with Hund's
interaction. In this way this quantum dot can be seen as a model system
for a Hund's coupled mixed-valence quantum impurity as appearing in
Hund's metals where local ferromagnetic interactions between orbitals
lead to the emergence of complex electronic states.
\end{abstract}
\maketitle
Electronic transport measurements in mesoscopic systems sometimes
show a fascinating phenomenon, a peak or a dip at zero bias, a so-called
zero bias anomaly (ZBA) \cite{Logan1964,Wyatt1964,Zeller1969}. The interest in this anomaly originates from
the fact that it indicates emergent phenomena in the systems under
consideration not captured by traditional (semi-classical) transport
theories. In superconducting systems they can appear due to Andreev
reflections or due to bound states at interfaces. The most prominent
examples are the Majorana zero modes in topological superconductor
structures which can show the presence of non-Abelian anyons and which
may open up the possibility for topological quantum computing \cite{Kitaev2003,Oreg2010,Lutchyn2010}.
In such systems ZBAs were observed \cite{Mourik2012}, but subsequently
quite controversially discussed. Very recently a publication claimed
to show first steps into topological quantum computing with Majorana
zero modes \cite{microsoft2025}. ZBAs can also appear in disordered
systems because of anomalies in the density of states due to electron-electron
interactions \cite{Altshuler1979}. One of the most fascinating ZBAs
originates from the Kondo effect, the interaction of conduction electrons
with localized magnetic moments \cite{Kondo1964}. The characteristic
feature of the highly-entangled Kondo state is a many-body resonance
in the spectral function that forms at low-temperatures in the vicinity
of the chemical potential. In quantum dots (QD) it gives rise to a
ZBA \cite{Glazman1988,Ng1988,Meir1992, Koenig1996} and this increased conductance at zero
bias typically appears for odd numbers of electrons on the dot \cite{Goldhaber-Gordon1998,Cronenwett1998,Schmid1998,Nygard2000}.
Then a spin singlet is formed between the localized electron spin
on the dot and the screening cloud of electrons in the leads\cite{Heine2016,V.Borzenets2020}.
For small numbers of electrons on a QD, the appearance of a shell like
structure was shown and consecutive filling according to Hund's rule
was observed \cite{Kouwenhoven2001}. In multi-orbital systems, as
e.g. in diluted magnetic alloys, the interplay between Hund's rule
coupling and Kondo effect has gained high attention during the past
decade \cite{Nevidomsdyy2009, Florens2011, Medici2011, Stadler2015, Deng2019, Drouin2022}. The interest has been rekindled in connection with ``Hund's
metals'' \cite{Georges2024} and transition metal adatoms on metallic
surfaces \cite{khajetoorians2015}.

The present letter further explores the interplay of Hund's rule coupling
and Kondo effect. The high tunability of QDs allows to extend the
studies under conditions not accessible for dilute magnetic alloys.
Central for the present investigations is the ability to precisely
count the number of electrons on the QD and to vary the occupation
in a controlled way. We are concentrating our studies on the third
shell of the QD which is the lowest lying shell with different local
multiplets for different fillings  and we find ZBAs with unexpected
widths originating from Kondo effect and Hund's multiplets. \\

\begin{figure*}[t]
\centerline{\includegraphics[scale=0.9]{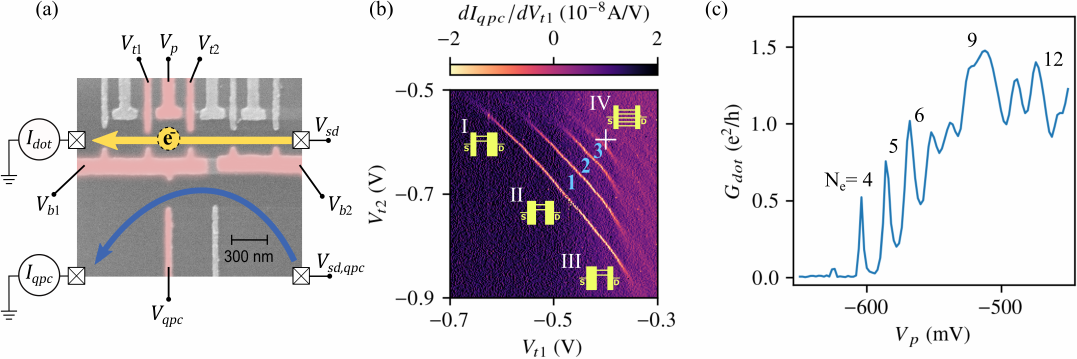}}
\caption{ (a) SEM image of the device structure with the schematic experimental
set-up. The quantum dot is described by the yellow $e^{-}$ island
and formed electrostatically by applying negative potentials. The
QPC is defined by $V_{qpc}$. The gray gates are not used. The crossed
squares indicate the ohmic contacts of the sample. (b) Charge stability
diagram of the quantum dot in the few-electron regime as a function
of the tunnel barrier gate voltages $V_{t1}$ and $V_{t2}$. The visible charging
line below the indicated $N_{e}=1$ is the charging line of the first electron.
The symmetry of tunnel coupling can be adjusted along the charging lines.
In region II, about the middle of the visible
charging line, the tunnel barriers are symmetrical. For more positive tunnel barrier gate
voltage $V_{t1}$ or $V_{t2}$, the tunnel barriers become asymmetrical
(Region I und III).  The more positive the tunnel barrier gate voltages
are, the stronger the tunnel coupling becomes (Region IV). (c) The
conductance $G_{dot}$ as a function of the plunger gate voltage $V_{p}$.
}
\label{fig:1} 
\end{figure*}

For our studies we use a Schottky gate defined
quantum dot based on a two-dimensional electron gas (2DEG) in a standard
GaAs/AlGaAs heterostructure. The $10$ $nm$ thick 2DEG is formed
approximately $100$ $nm$ below the surface with an electron density
of $n_{e}=2.4$ x $10^{11}$ $cm^{-2}$ and a mobility of $\mu_{e}=5.1$
x $10^{5}$ $cm^{2}V^{-1}s^{-1}$. The device and its characteristic
properties are displayed in Figure \ref{fig:1}. A scanning electron
microscope (SEM) image of the device structure is shown in Fig. \ref{fig:1}(a).
The structure would allow for the definition of up to four dots, but
in the work here only one dot was used. The quantum dot ($e^{-}$
island) is formed electrostatically by applying negative voltages
to the two tunnel barrier gates $t1$ and $t2$, the plunger gate
$p$ and a larger gate $b1$. The quantum dot can be charged or discharged
via two tunnel-coupled electron reservoirs. The coupling to the reservoirs
can be manipulated via the tunnel barrier gates $t1$ and $t2$, while
energy levels can be controlled via gate $p$. The single-electron
charging process is sensed directly with a capacitively coupled quantum
point contact (QPC) similar to the one used in Ref.\cite{Wagner2017}.
The QPC is formed by the gates $b1,b2$, and $qpc$ and used as a
charge detector which is mainly defined by the voltage $V_{qpc}$.
The gap between gates $b1$ and $b2$ is electrostatically closed
by applying sufficiently negative potentials, separating the quantum
dot circuit from the detector circuit. The measurements are carried
out at a base temperature of $\SI{10}{mK}$. 

 Figure \ref{fig:1}(b) presents a gate-dependent charge stability
diagram of the quantum dot in the few-electron regime. The detector
signal is shown as function of the tunnel barrier gate voltages $V_{t1}$
and $V_{t2}$ for a plunger gate voltage $V_{p}=\SI{-600}{mV}$. It is given
by the numerical derivative of the detector current $dI_{qpc}/dV_{t1}$.
The first visible charging line is the charging line of the first
electron to the quantum dot $N_{e}=1$. Each charging line indicates
an increase in the number of electrons in the quantum dot by one.
In this way we know the exact number of electrons occupying the quantum
dot for a given plunger gate voltage.
The coupling symmetry and the strength can be adjusted using the stability diagram, as demonstrated in Figure \ref{fig:1}(b).
 The Kondo effect in quantum dots occurs typically for symmetric and
strong tunnel couplings to the electron reservoirs \cite{Goldhaber-Gordon1998,Cronenwett1998,Schmid1998,Nygard2000,Heine2016,simmel1999}.
Therefore, to investigate Kondo effect in our structure we have set
the tunnel gate voltages to $V_{t1}=\SI{-400}{mV}$ and $V_{t2}=\SI{-595}{mV}$ (white cross in Fig. \ref{fig:1}(b)), i.e. at quasi-symmetric
barriers. For these barrier gate voltages the conductance $G_{dot}$
through the QD is then measured as a function of the plunger gate
voltage $V_{p}$ (Fig. \ref{fig:1}(c)). From the measurements shown
in Fig. \ref{fig:1} (b), it is possible to determine the number of
electrons $N_{e}$ occupying the QD, as shown in Fig. \ref{fig:1}(c). The conductance of the dot shows
a strong Coulomb blockade for $N_{e}=4,5,6$ and $12$, whereas the
minima in between corresponding to electron numbers $N_{e}=7$ up
to $N_{e}=11$ are much weaker. Such strong Coulomb blockade at $N_{e}=6$
and $N_{e}=12$ and weaker effects in between are clear signs that
we observe here the filling of the third shell of the QD confined
within the 2DEG of our heterostructure \cite{Tarucha1996,Kouwenhoven2001}.
Interestingly for the electron number of $N_{e}=9$ we observe in
Fig. \ref{fig:1}(c) not a real minimum in the conductance, but more
or less absence of Coulomb blockade. 
\begin{figure}[t]
\centerline{\includegraphics[scale=0.825]{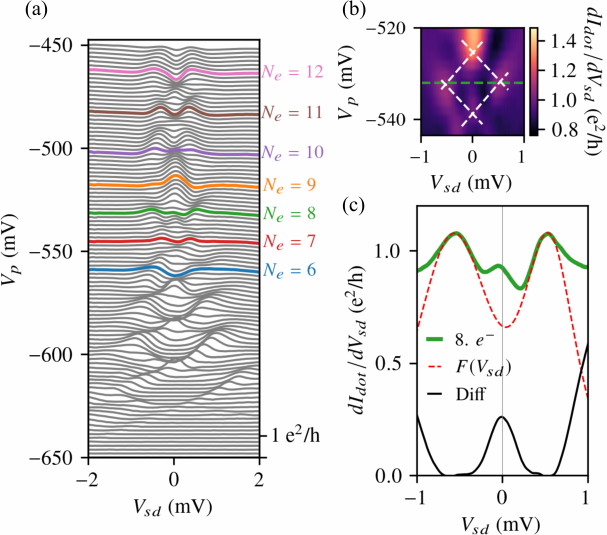}}
\caption{ (a) Waterfall plot of the Coulomb blockade diamonds differential
conductance $dI_{dot}/dV_{sd}$. The colored lines
are linecuts through the Coulomb diamonds for $N_{e}$ electrons occupying
the quantum dot. (b) The measured $dI_{dot}/dV_{sd}$
as a function of the plunger gate voltage $V_{p}$ and the bias
voltage $V_{sd}$. The dashed white lines highlight the Coulomb diamond
for $N_{e}=8$. (c) Analysis of the differential conductance for $N_{e}=8$. The
green line is section of the conductance along the dashed green line
in (b) at a gate plunger gate voltage $V_{p}=\SI{-532}{mV}$. The Coulomb
resonance peaks are fitted with the function $F(V_{sd})$ (dashed
red line) and subtracted from the measured conductivity, resulting
in the black line.  }
\label{fig:2} 
\end{figure}
 To further characterize the electronic structure of the QD, we investigated
the differential conductance ${dI_{dot}}/{dV_{sd}}$ as function of
applied source-drain bias $V_{sd}$ for a certain gate voltage range
covering the filling of the third shell as shown in the waterfall
plot in Fig. \ref{fig:2}(a). The results displayed in Figure \ref{fig:1}(c)
correspond to the zero-bias line $V_{sd}=0$. The traces of the measurements
for the electron numbers $N_{e}=6$ to $N_{e}=12$ are highlighted with
colored lines. In these colored lines, clear minima are observed
for electron numbers 6 and 12 at zero bias. For $N_{e}=9$ the variation
with bias $V_{sd}$ of the differential conductance exhibits a pronounced
maximum at zero bias. For electron numbers 7, 8, 10, and 11 the situation
is less clear. Let us consider the case $N_{e}=8$ in greater detail.
Figure \ref{fig:2}(b) shows the differential conductance in a color
scale for the Coulomb diamond of $N_{e}=8$. The dashed green line marks
the middle of the diamond at a plunger gate voltage of $V_{p}=\SI{-532}{mV}$.
The green line in Fig. \ref{fig:2}(c) gives the differential conductance
as function of bias voltage $V_{sd}$ at this plunger gate voltage.
For these $N_{e}=8$ electrons one sees
a certain maximum in the differential conductance at zero bias, i.e.
a ZBA, as well as two strong maxima at $V_{sd}=\SI{0.523}{mV}$ and $\SI{-0.554}{mV}$
originating from Coulomb resonances in Figure \ref{fig:2}(b). For
samples with small charging energies and strong coupling to the leads,
as in our case here, the Coulomb resonance peaks overlap within the
Coulomb-blocked region resulting in an increased conductivity inside
the Coulomb blockade region. This additional conductivity has to be
subtracted for a more detailed analysis of the ZBA. The result is
displayed in Fig. \ref{fig:2} (c). The Coulomb peaks can be fitted
with a function $F\left(V_{sd}\right)$ (see Supplemental Material) \cite{Beenakker1991}. The dashed red line in Fig. \ref{fig:2}(c)
is the result of the best fitting of the two Coulomb resonances with
the function $F\left(V_{sd}\right)$. By subtracting this fitted curve
from the measured conductivity, we obtain the differential conductivity
without the contribution of the Coulomb resonances (black line Fig.
\ref{fig:2}(c)). A clear maximum is seen at zero bias. 
This ZBA could hint towards
Kondo effect, but in this case for an even number of electrons ($N_{e}=8$).
In a small number of previous experiments deviations of the typical
odd-even behaviour for the Kondo effect were observed and ZBAs were
found for consecutive numbers of electrons, but details were not clear
\cite{Schmid2000,Fuhner2002,Keyser2003}.
The black curve in Fig. \ref{fig:2}(c) shows in addition to
the ZBA increasing differential conductivities in approaching $V_{sd}=-1$
$mV$ and $+1$ $mV$ that could originate from contributions of excited
states in the dot and also from co-tunneling processes due to the
strong coupling of the quantum dot to the leads. The procedure of
fitting and subtracting the charging peaks was applied to all the
measurements highlighted in color in Fig. \ref{fig:2}(a). The measured
differential conductivities for $N_{e}=7-11$ are plotted in Fig.
\ref{fig:3}(a) with the $F\left(V_{sd}\right)$ fits of the two Coulomb
resonances for each $N_{e}$ (dashed red lines). Twice the addition
energy $E_{c}$ can be obtained from the difference in the Coulomb
peak positions. The addition energy $E_{c}$ is given by $E_{c}(N_{e})=\mu_{dot}(N_{e}+1)-\mu_{dot}(N_{e})=E(N_{e}+1)-2E(N_{e})+E(N_{e}-1)$
where $\mu_{dot}$ is the chemical potential of the dot and $E$ the
ground state energy of the dot \cite{Kouwenhoven2001}. The successive
filling of the third shell of a QD is schematically shown in Fig.
\ref{fig:3}(b) for the electron numbers 6 to 9, i.e. up to half filling
of the third shell whereas the first and second shell are filled and
not changed. Figure \ref{fig:3}(c) presents $E_{c}$ for $N_{e}=6$
to $N_{e}=12$ as extracted from our experiment. Since $E_{c}$ can
be defined as a difference in chemical potentials, the shown addition
energies are related to the addition of $7$th to $13$th electron
to the QD. $E_{c}$ for $N_{e}=6$ corresponds to adding the first
electron in the third shell, while $E_{c}$ at $N_{e}=12$ corresponds
to the first electron in the next shell \cite{Tarucha1996,Rontani1998}.
The observed decrease in $E_{c}$ for $N_{e}=7$ is due to the fact
that the same shell is filled. A triangular evolution of $E_{c}$
is observed for $N_{e}=7-11$ with a peak at half-shell filling.
Here, $E_{c}$ for $N_{e}=9$ is even higher than the
addition energy at $N_{e}=6$ and at $N_{e}=12$ electron, in contrast to Ref. \cite{Tarucha1996,Rontani1998}. 
\begin{figure}[t]
\centerline{\includegraphics[scale=0.9]{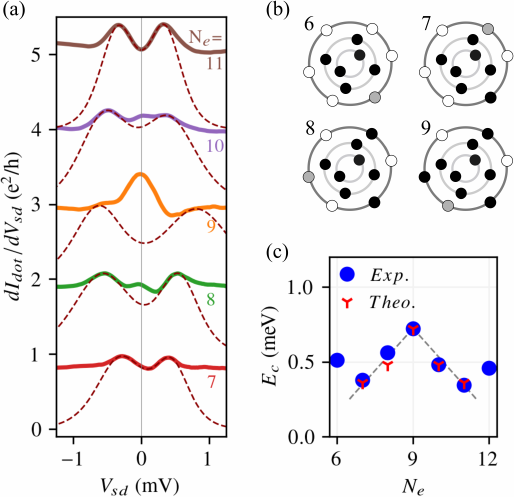}}
\caption{ (a) The measured differential conductance as a function of the bias
voltage $V_{sd}$ for the different occupation numbers $N_{e}$. The
dashed red lines show the $F(V_{sd})$ fits of the Coulomb peaks.
The signal is offset ($+1$ $e^{2}/h$) to increase the visibility.
(b) Schematic representation of the addition of electrons in the third
shell. The gray dots indicate the addition of the next electron. (c) The addition energy as a function of $N_{e}$. The shown theoretical values
are calculated for $U=0.36$ $meV$ and $J=0.18$ $meV$. The dashed
gray lines are guide lines showing the triangular behavior of the addition
energies. }
\label{fig:3} 
\end{figure}

To model our results, we consider the complex dynamics of the dot electrons in a partially
occupied shell with three-fold orbital degeneracy while the electrons
of the filled first and second shell are accounted for by the contribution
of their densities to the confining electrostatic potential. The electrons
occupying the six single-particle levels interact via their intra- and inter-orbital Coulomb interactions $U$
and $U'$ as well as the ferromagnetic Hund's rule exchange coupling
$J$. The central focus of our study is the role of the exchange interaction.
 For simplicity we assume $U=U'$, i. e., we adopt a Constant Interaction
model augmented by Hund exchange 
\footnote{The present parametrization differs from the commonly used Kanamori
parametrization which is applicable for transition metal atoms in
a cubic environment where the d-states are split $t_{2g}$ and $e_{g}$
by the Crystal Field} 
as has been successfully applied
to QDs (see \cite{Kouwenhoven2001} and references therein) (see End Matter).  

To qualitatively understand the experimental findings we adopt Linear
Response Theory which relates the variation with source drain voltage
$V_{sd}$ of the differential conductance to the dot spectral function
$\rho_{\nu\sigma}\left(\omega\right)$ measuring the probability for adding
or removing a dot electron in orbital $\nu\sigma$ with energy $\omega$.
The low-temperature spectral function of an isolated QD exhibits charge
peaks shifted and split by the exchange interaction. The addition
energies are easily calculated yielding $E_{c}(N_{e})=U,U+\frac{2}{3}J,U+2J,U+\frac{2}{3}J,U$
for $N_{e}=7,8,9,10,11$, respectively (see supplemental material), which nicely explain the triangular
variation with $N_{e}$ of the addition energies in Figure \ref{fig:3}(c)
as direct consequence of the exchange interaction. 

\begin{figure}[h]
\centerline{\includegraphics[scale=0.9]{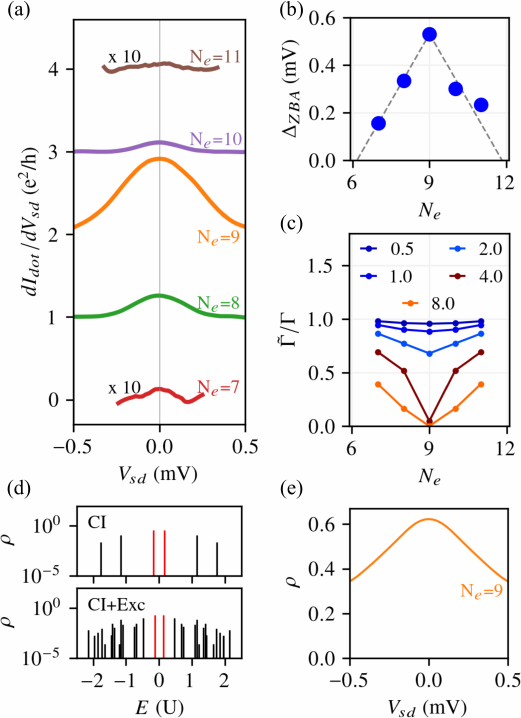}}
\caption{(a) The differential conductivity after subtracting $F(V_{sd})$.
The signal is offset ($+1$ $e^{2}/h$) to increase the visibility.
(b) The width of the ZBA for the different occupation numbers $N_{e}$.
The dashed lines are guides showing the triangle behaviour of the
width. (c) Renormalization of the Kondo resonance widths $\tilde{\Gamma}/\Gamma$:
Variation with $N_{e}$ as predicted by RISB for fixed ratio $J/U=0.3$
and various correlation strengths $U/\pi\Gamma$. (d) The spectral function for Zero Band Width model
for $N_{e}=9$ and $0.2U$. The upper panel is for $J=0$ (CI), while the lower panel for $J=0.3U$ (CI+Exc). (e) Broadened spectral function according to the model shown in lower panel of (d).
}
\label{fig:4} 
\end{figure}


In the following we come back to the experimentally observed ZBAs,
i.e. the interplay between Kondo effect and Hund's rule correlations
for the filling of the third shell. The resulting conductivities from
subtracting charging peaks in the experimental data as shown in Fig.
\ref{fig:3}(a) are plotted in Fig. \ref{fig:4}(a) as function of
$V_{sd}$ for $N_{e}=7-11$. Clearly, for all these numbers of electrons ZBAs are observed, but without an even-odd effect. Instead there seems
to be a certain symmetry and scaling hidden in the strengths of the
ZBAs as function of the filling of the shell. Whereas for the first electron
in the shell $N_{e}=7$ and for the last missing electron in the shell
$N_{e}=11$ only very small and noisy maxima of about $0.01e^{2}/h$
are obtained, for $N_{e}=8$ and $N_{e}=10$ the maxima have amplitudes
of about $0.1e^{2}/h$ and for the half filling of the shell at $N_{e}=9$
even a maximal value of the amplitude of the ZBA on the order of $e^{2}/h$
is observed. These observations show a certain electron-hole symmetry
for this shell which originates from the filling of the three spin-degenerate
orbitals. Figure \ref{fig:4}(b) shows the width $\Delta_{ZBA}$ 
of the ZBAs plotted as function of electron number $N_{e}$.
We see again a maximum at an electron number at half filling of the third
shell of our quantum dot. Going away from $N_{e}=9$, $\Delta_{ZBA}$ decreases more or less linearly, as indicated by the dashed
lines. It seems to show a triangle like behaviour in this shell,
indicating again an electron-hole symmetry very similar to the addition
energies as discussed in Fig. \ref{fig:3}(c). We already understood
that the triangle like behaviour for the addition energies is connected
to the exchange interaction via Hund's rule. Therefore, in order to
understand the observation of the triangle-like behaviour for the
width of the ZBAs we turn in the following to the interplay between
multiplet formation via Hund's rule correlations (see supplemental
material) and momentum screening via the Kondo effect.

For the Kondo effect we study the variation with $N_{e}$ of the effective
width $\tilde{\Gamma}$ and position $\tilde{\epsilon_{d}}$ of the
many-body resonance that appears in the dot spectral function at low
temperatures. These parameters can be calculated within Slave Boson
Mean Field (SBMF) theory which provides a very useful tool in treating
the low-energy properties of crystalline materials with strongly correlated
electrons. To account for spin correlations, we employ the multi-orbital
Rotationally-Invariant Slave-Boson (RISB) (see \cite{Fresard15} for
a review and references therein). In the supplement, we briefly summarize
the method and show that our RISB ansatz reproduces the NRG results
for a two-orbital Anderson model with Hund's rule correlation at integer
occupancies provided we replace the hybridization strength $\Gamma_{NRG}\to2\Gamma_{RISB}$
as suggested by \cite{Kleeorin2018}. For the three-orbital Anderson
model, the variation with $N_{e}$ of the Kondo resonance width in
Figure \ref{fig:4}(c) clearly exhibits a pronounced minimum at half-filling
independent of correlation strength and in agreement with the expected
exponential sharpening of the Kondo resonance \cite{Schrieffer1967, Daybell1968}. This behaviour is in
marked contrast to the variation of the ZBA widths in Fig. \ref{fig:4}(b).
This suggests that the observed ZBAs do not exclusively result from
the quasiparticle Kondo resonances.

To further analyse the origin of the ZBAs we replace the states in
the leads by discrete levels with energy $\epsilon_{c}$. The spectral
function of this generalized Zero Band Width Anderson Model or Molecular
Model \cite{Fulde1988,Runge1996,HewsonBook} at half-filling is shown
for a pure constant interaction model (CI) in the upper half of Fig. \ref{fig:4}(d)
and for a model involving also exchange(CI+Exc) in the lower half
of Fig. \ref{fig:4}(d). The spectral function for the constant interaction
model (CI) without Hund exchange shows resonances for charge states
and for Kondo. There is a transfer of spectral weight to the low-energy
regime around the Fermi energy highlighting the Kondo effect (Kondo
resonances are shown in red in the spectral function). Such a simple
model would not be able to explain our broad ZBAs as discussed above.
In the spectral function for a model including exchange (CI+Exc) a
multitude of resonances are appearing. In addition to the transfer
of spectral weight to the low-energy regime around the Fermi energy
highlighting the Kondo effect, there are features observed at finite
energies due to the existence of excited Hund multiplets. Due to the
threefold orbital degeneracy, a multitude of excited multiplets are
possible (see Supplemental material). For the total spin of the system,
we observe in our Zero-bandwidth model (see Endmatter) that it vanishes
and quenches at low temperatures showing the perfect screening of
the spin and proving the presence of Kondo effect. Therefore, one can
argue that the amplitude of the experimentally observed ZBAs originates
from the Kondo effect whereas the width is influenced by excited Hund
multiplets. In order to compare our calculated spectral function with
the experimental observation we assume a finite band width. Figure
\ref{fig:4}(e) shows the result of this broadened spectral function
which is nicely comparable to the experimentally obtained ZBA as shown
for $N_{e}=9$ in Fig. \ref{fig:4}(a). This rather broad low-energy
feature at half-filling indicates the existence of the above mentioned
satellites from excited multiplets, in close analogy to the well-known
spin-orbit satellites in the Kondo resonance of heavy-fermion materials \cite{Bickers1987}.
Due to the existence of these satellite features, it is not obvious
how one can deduce a characteristic Kondo temperature from the observed
ZBA. Nevertheless, as shown in EndMatter, the ZBA vanishes with increasing
temperature being consistent with the expectations of a Kondo effect.
Interestingly the apparent addition energy $E_{c}$ is slightly decreasing
with increasing temperature due to this vanishing of the Kondo effect
and the removing of weight from the Kondo resonance.
 
In conclusion, ZBAs are observed for successive filling of the
third shell of a QD. There is no even odd effect showing up, but a
characteristic particle-hole symmetry for the three spin-degenerate
orbital states. 
These observed broad ZBAs are explained by the interplay of Hund's rule and Kondo
effect and by the presence of excited Hund multiplets. In this way
our quantum dot serves as a model system for a Hund's coupled impurity.

\begin{acknowledgments}
This work was supported by the Deutsche Forschungsgemeinschaft (DFG,
German Research Foundation) under Germany's Excellence Strategy -
EXC 2123 QuantumFrontiers - 390837967 and The State of Lower Saxony
of Germany via the Hannover School for Nanotechnology. 
\end{acknowledgments}

\appendix

\makeatletter \providecommand{\@ifxundefined}[1]{%
	 \@ifx{#1\undefined}
	}\providecommand{\@ifnum}[1]{%
	 \ifnum #1\expandafter \@firstoftwo
	 \else \expandafter \@secondoftwo
	 \fi
	}\providecommand{\@ifx}[1]{%
	 \ifx #1\expandafter \@firstoftwo
	 \else \expandafter \@secondoftwo
	 \fi
	}\providecommand{\natexlab}[1]{#1}\providecommand{\enquote}[1]{``#1''}\providecommand{\bibnamefont}[1]{#1}\providecommand{\bibfnamefont}[1]{#1}\providecommand{\citenamefont}[1]{#1}\providecommand{\href@noop}[0]{\@secondoftwo}\providecommand{\href}[0]{\begingroup \@sanitize@url \@href}\providecommand{\@href}[1]{\@@startlink{#1}\@@href}\providecommand{\@@href}[1]{\endgroup#1\@@endlink}\providecommand{\@sanitize@url}[0]{\catcode `\\12\catcode `\$12\catcode
	  `\&12\catcode `\#12\catcode `\^12\catcode `\_12\catcode `\%12\relax}\providecommand{\@@startlink}[1]{}\providecommand{\@@endlink}[0]{}\providecommand{\url}[0]{\begingroup\@sanitize@url \@url }\providecommand{\@url}[1]{\endgroup\@href {#1}{\urlprefix }}\providecommand{\urlprefix}[0]{URL }\providecommand{\Eprint}[0]{\href }\providecommand{\doibase}[0]{https://doi.org/}\providecommand{\selectlanguage}[0]{\@gobble}\providecommand{\bibinfo}[0]{\@secondoftwo}\providecommand{\bibfield}[0]{\@secondoftwo}\providecommand{\translation}[1]{[#1]}\providecommand{\BibitemOpen}[0]{}\providecommand{\bibitemStop}[0]{}\providecommand{\bibitemNoStop}[0]{.\EOS\space}\providecommand{\EOS}[0]{\spacefactor3000\relax}\providecommand{\BibitemShut}[1]{\csname bibitem#1\endcsname}\let\auto@bib@innerbib\@empty

\section*{End Matter}

\sect{Model Hamiltonian and calculational methods} In order to analyse
the observed triangular evolution of $E_{c}$ in more detail we model
the QD coupled to leads by a multi-orbital Anderson model 
\begin{equation}
H=H_{dot}+H_{hyb}+H_{leads}
\end{equation}
where
\begin{eqnarray}
H_{dot} &= &\sum_{\nu\sigma}\left[\epsilon_{d}n_{d\nu\sigma}+\frac{1}{2}Un_{d\nu\sigma}n_{d\nu-\sigma}\right]\nonumber\\
 & &
+\frac{1}{2}\text{\ensuremath{\sum}}_{\nu\neq\nu',\sigma\sigma'}\left[U'n_{d\nu\sigma}n_{d\nu'\sigma'}-2J{\bf S}_{d\nu}\cdot{\bf S}_{d\nu'}\right]
\end{eqnarray}
describes the (strongly) correlated electrons of the QD. While
\begin{eqnarray}
H_{hyb} & = &\sum_{k\nu\sigma}\mathcal{\mathscr{V}}_{\nu}\left(c_{k\nu\sigma}^{\dagger}d_{\nu\sigma}+d_{\nu\sigma}^{\dagger}c_{k\nu\sigma}\right)
\end{eqnarray}
models the weak coupling of the latter to the 2DEG in the leads accounted
for by
\begin{eqnarray}
H_{leads} & = & \sum_{k \nu \sigma}\epsilon_{k}c_{k\nu\sigma}^{\dagger}c_{k\nu\sigma} \quad .
\end{eqnarray}

Here $c_{k\nu\sigma}^{\dagger}(c_{k\nu\sigma})$ and $d_{\nu\sigma}^{\dagger}(d_{\nu\sigma})$
denote creation (annihilation) operators for an electron in the leads
with energy $\epsilon_{k}$, spin $\sigma$, and
orbital symmetry $\nu$ and one with spin $\sigma$
in orbital $\nu$ on the QD, respectively. For the latter, the number
operators $n_{d\nu\sigma}=d_{\nu\sigma}^{\dagger}d_{\nu\sigma}$ and
the components of the spin density operators ${\bf S}_{d\nu}$ are
given by $S_{d\nu}^{i}=\frac{1}{2}d_{\nu\sigma}^{\dagger}\bm{\tau}_{\sigma\sigma'}^{i}d_{\nu\sigma'}$
with the usual Pauli matrices $\bm{\tau}^{i}$, $i=1,2,3$. Throughout
the calculations we measure energies relative to the Fermi energy
in the leads and use $\hbar=1$. Each orbital $\nu$ of the QD is
individually coupled to an autonomous conduction electron bath formed
by particular linear combinations of source and drain. As the number
of screening channels is sufficient to fully compensate the total
spin on the QD, the ground state is a spin singlet giving rise to
the usual Fermi liquid behaviour similar to the single-channel Kondo
model \cite{Nozieres1980,Schlottmann1993}. In the present studies
we neglect the anisotropies in the spin-conserving hybridization setting
$\mathscr{V}_{\nu}=\mathscr{V}$ and define the hybridization strength
$\Gamma=\pi\mathscr{V}^{2}D(0)$ with the Density of States in the
leads $D\left(0\right)$.

\begin{figure}
\centerline{\includegraphics[width=1\columnwidth]{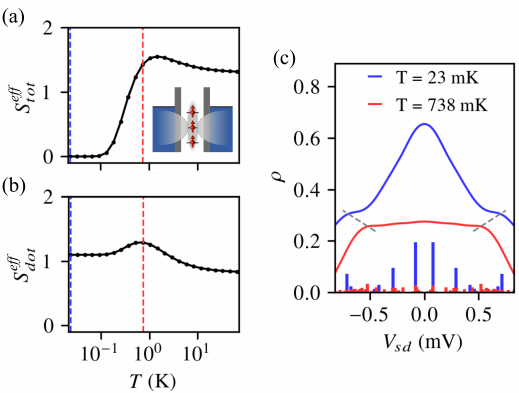}}
\caption{(a), (b) The total and dot spin as a function of temperature. (c)
The spectral function for Zero Band Width model along with the broadened
spectral function, $V_{sd}$ scaled, for $N_{e}=9$ at two different
temperatures. The spectral function is represented in blue for a low
temperature of $\SI{23}{mK}$ and in red for the temperature at which
the dot spin is maximum. The broadening width is around $\SI{0.45}{meV}$. }
\label{fig:I} 
\end{figure}

 \sect{Temperature dependences} The interplay between
Hund's rule and Kondo effect can be seen from the variation with temperature
of the effective total spin $S_{tot}^{eff}\left(T\right)$ and the
effective spin on the QD $S_{dot}^{eff}\left(T\right)$. The effective
total spin $S_{tot}^{eff}\left(T\right)$ is given by $S_{tot}^{eff}\left(T\right)\left(S_{tot}^{eff}\left(T\right)+1\right)=\left\langle \left({\bf S}_{d}+{\bf S}_{c}\right)^{2}\right\rangle _{T}$
whereas the effective spin on the QD $S_{dot}^{eff}\left(T\right)$
is given by $S_{d}^{eff}\left(T\right)\left(S_{d}^{eff}\left(T\right)+1\right)=\left\langle {\bf S}_{d}^{2}\right\rangle _{T}$
where $\left\langle O\right\rangle _{T}$ denotes the thermal average
$Tr\left\{ Oe^{-H/k_{BT}}\right\} /Tr\left\{ e^{-H/k_{BT}}\right\} $.
The results for the effective total spin $S_{tot}^{eff}\left(T\right)$
as function of temperature in a logarithmic scale are shown in Fig.
\ref{fig:I}(a), whereas for the effective spin on the QD $S_{dot}^{eff}\left(T\right)$
in Fig. \ref{fig:I}(b). Both quantities exhibit a maximum for temperatures
of the order of the exchange coupling $J$. The values are considerably
enhanced due to Hund's rule interaction. At low temperatures $T\to0$,
the effective total spin vanishes in the non-magnetic ground state
being a clear indication for Kondo effect. The pronounced Kondo feature
in the spectral function appears at temperatures where the total spin
is quenched.

For two characteristic temperatures the spectral function was calculated
as shown in the bottom of Fig. \ref{fig:I}(c.). At the low temperature
of $T=\SI{23}{mK}$ the spectral weight of the Kondo resonances is
clearly dominating the spectral function. In contrast at the higher
temperature the picture changes drastically. The temperature of $T=\SI{738}{mK}$
is given by the exchange coupling $J$ and corresponds to the temperature
of the maxima as observed in the effective spins shown in Figs. \ref{fig:I}(a)
and (b). For this high temperature the spectral weight of the Kondo
resonances is greatly reduced and instead the excited Hund multiplets
have gained weight. In order to compare these calculated spectral
weights to experimental results they are also shown in Fig. \ref{fig:I}(c.)
as function of source-drain voltage with a broadening of $\SI{0.45}{meV}$
applied (blue and red curves). At the low temperature (blue line)
one sees a broad ZBA as discussed before and in addition small shoulders
to the left and right originating from the charge states. The difference
in the position of these two shoulders is given by twice the addition
energy $E_{c}$ (as mentioned above). Interestingly for the higher
temperature the ZBA vanishes and instead an almost featureless spectral
function being finite over a wide range is observed with the shoulders
being located at about $\SI{-0.5}{mV}$ and at $\SI{0.5}{mV}$ (red
curve in Fig. \ref{fig:I}(c.)). The two shoulders are again characterizing
the addition energy $E_{c}$ which is clearly reduced in comparison
to the low temperature value (shown by the two dashed grey lines).
So, we see in these broadened spectral functions how the weight transfer
to the Kondo resonance increases the effective addition energy at
low temperatures whereas at high temperature this weight transfer
is reduced and the addition energy approaches the charging energy
as describable in a Constant interaction model.

\begin{figure}[h]
\centerline{\includegraphics[width=1\columnwidth]{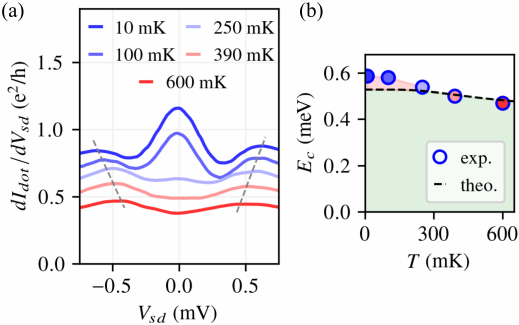}}
\caption{(a) The measured differential conductance as a function of the bias
voltage for $N_{e}=9$ at temperatures ranging from $T=\SI{10}{mK}$
to $\SI{600}{mK}$. The signal is offset ($\SI{-0.1}{e^{2}/h}$) to increase
the visibility. The dashed gray lines are guide lines for the Coulomb
peak positions. (b) The charging energy for the different temperature.
The dashed black line is a fit using Eq. (\ref{eq:E_c}). }
\label{fig:II} 
\end{figure}

In order to compare these calculated results with experimental data,
measurements for the QD are shown in Figure \ref{fig:II}(a) and (b)
(in a slightly different situation than presented in Figs. \ref{fig:3}
and \ref{fig:4}, with the tunnel gate voltages set to $V_{t1}=\SI{-425}{mV}$
and $V_{t2}=\SI{-630}{mV}$). Figure \ref{fig:II}(a) shows the differential
conductance as a function of the source-drain voltage $V_{sd}$ for
temperatures ranging from $\SI{10}{mK}$ to $\SI{600}{mK}$. The gate
voltage $V_{p}$ is set at the center of the Coulomb diamond for $N_{e}=9$.
With increasing temperatures the broad ZBA vanishes as expected for
Kondo effect. Interestingly at the same time the two Coulomb resonances
approach one another, as indicated by the dashed gray lines (serving
as guidelines for the Coulomb peak positions). The peak-to-peak distance
of the Coulomb peaks divided by two gives a measure for the addition
energy $E_{c}$. The results for $E_{c}$ are shown as small colored
circles in Fig. \ref{fig:II}(b) (colors correspond to the colors
and temperatures as shown in Fig. \ref{fig:II}(a). As the temperature
is raised, the Kondo effect vanishes and in addition, the excited Hund
multiplets become thermally populated leading to the observed changes
in the addition energies. In order to distinguish between the two
effects we used a simple model of calculating the effect of temperature
for an isolated dot with threefold degenerate orbital states and the
resulting Hund's multiplets (see Supplemental material). As the temperature
is raised, the excited Hund's multiplets will be thermally populated.
These changes in population of states affect also the addition energy
and reduce it for higher temperatures. For exchange energy $J$ being
smaller than charging energy $U$, one obtains in the simple model
of an isolated QD for the addition energy $E_{c}$ the equation 
\begin{equation}
E_{c}=U+2J\dfrac{1-e^{-3J/k_{B}T}}{1+3e^{-3J/2k_{B}T}+e^{-3J/k_{B}T}}\label{eq:E_c}
\end{equation}
The dashed line in Fig. \ref{fig:II}(b) (marked as $theo$) shows
the temperature dependence of $E_{c}$ according to this equation,
with $U=\SI{0.41}{meV}$ and $J=\SI{0.06}{meV}$. One sees that the small reduction at
the higher temperatures is nicely described by this formula for an
isolated dot, whereas at lower temperatures the appearance of the
Kondo effect leads to a stronger increase of the addition energies
$E_{c}$ with decreasing temperatures due to the shifting of the according
weights in the spectral function. This additonal increase due to Kondo
effect is colored in red in Fig. \ref{fig:II}(b). \\
 
\section*{Supplemental Material}

	\section{Fit function of the Coulomb peaks}

\begin{figure}[h]
\centerline{\includegraphics[scale=0.9]{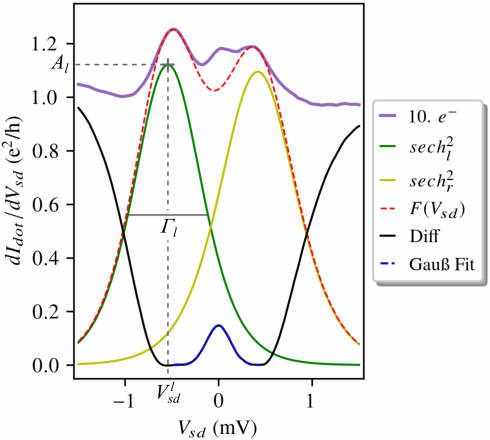}}
\caption{\label{fig.:BG}  The purple line is the conductivity as a function of source-drain voltage for a plunger gate voltage $V_{p}=\SI{-502}{mV}$. The $F(V_{sd})$ (dashed red line) is the summe of the two $sech^2$ functions, $sech^2_l$ and  $sech^2_r$ for the left and right peaks, respectively. The parameters $ A_l$, $V_{sd}^{l}$ and $\Gamma_l$ of $sech^2_l$ are highlighted, for reference. The difference of measured data and fitting function (black line) shows a ZBA, which is fitted with the Gaussian function. 
} 
\label{S_I}
\end{figure}

For more detailed examination of the ZBA, the additional conductivity caused by the overlap of the Coulomb resonances is subtracted. According to Ref. \cite{Beenakker1991}, the Coulomb peaks have each the form of a $sech^2$ function. Thus, they can be fitted as a function of source-drain voltage with the following function, as shown in figure \ref{S_I} for $N_e = 10$, 
\begin{eqnarray}
F(V_{sd})=A_{l} sech^{2}\left(1.76 \frac{V_{sd}-V_{sd}^{l}}{\Gamma_{l}}\right) \nonumber\\
   +A_{r} sech^{2}\left(1.76 \frac{V_{sd}-V_{sd}^{r}}{\Gamma_{r}}\right)
\label{eq:one}
\end{eqnarray}
where $V_{sd}^{l}$ and $V_{sd}^{r}$ are the positions of the two Coulomb resonance peaks maxima to the left $(l)$ and right $(r)$ of the zero point ($V_{eff, sd} =\SI{0}{mV}$ ). $ A_l$ and $A_r$ are the amplitudes of the left and right Coulomb resonances. 
By inserting the factor $1.76$, $\Gamma_l$ and $\Gamma_r$ give the full width at half the maximum (FWHM) of each the $sech^2$ functions. 

\section{Theory}
\subsection{Spectral function and addition energy of an isolated QD}

The current through the dot is given by 
\begin{align}
I\left(V_{sd}\right)\sim & \sum_{\nu\sigma}\int_{-\infty}^{+\infty}d\omega\left[f\left(\omega-\alpha eV_{sd}\right)-f\left(\omega+\left(1-\alpha\right)eV_{sd}\right)\right]\nonumber\\
 & t_{\nu\sigma}\left(\omega,V_{sd},T\right)
\end{align}
with the Fermi function $f\left(\omega\right)$, the difference in
chemical potential across the dot $eV_{sd}$ and the transmission
function $t_{\nu\sigma}\left(\omega,V_{sd},T\right)$. In the following
discussion, we adopt Linear Response theory and neglect orbital anisotropies
in the coupling to the leads. The transmission function is proportional
to the spectral function $\rho_{\nu\sigma}\left(\omega,T\right)$.
At low temperatures $T\to0$, the variation with source drain voltage
of the differential conductance 
\begin{equation}
\frac{dI}{dV_{sd}}\sim\sum_{\nu\sigma}\alpha\rho_{\nu\sigma}\left(\alpha eV_{sd}\right)+\left(1-\alpha\right)\rho_{\nu\sigma}\left(-\left(1-\alpha\right)eV_{sd}\right)
\end{equation}
is directly given by the energy dependence of the spectral function.
In addition, we assume $\alpha=1/2$.

The spectral function of an isolated QD consists of discrete lines
at energies $\epsilon_{i\left(N\right)}$ with weights $w_{i\left(N\right)}$
for the transitions $N\to N\pm1$ in a QD with $N$ electrons in the
ground state 
\begin{equation}
\rho_{\nu\sigma}\left(\omega\right)=\sum_{i\left(N\right)}w_{i\left(N\right)}\delta\left(\omega-\epsilon_{i\left(N\right)}\right)\quad,
\end{equation}
and, consequently 
\begin{equation}
\frac{dI}{dV_{sd}}\sim\sum_{i\left(N\right)}w_{i\left(N\right)}\left[\delta\left(V_{sd}-2\epsilon_{i\left(N\right)}\right)+\delta\left(V_{sd}+2\epsilon_{i\left(N\right)}\right)\right]\quad.
\end{equation}
The theory values, in Fig. 3(c) of the main text, are the centers of gravity of the exchange-split
charge peaks in the differential conductance given by the averages
$\bar{E}_{c}=2\sum_{i\left(N\right)}w_{i\left(N\right)}\left|\epsilon_{i\left(N\right)}\right|$.

Within our model, the energies of the many-electron states in the
partially filled third energy shell are labelled by the set of quantum
numbers $\left(N,S,\xi\right)$ where $N=N_{e}-6$ is the number of
electrons in the third shell, $S$ is the total spin, and $\xi$ is
the seniority, i. e., the number of unpaired electrons. The energetically
degenerate eigenstates corresponding to these levels are further characterized
by the spin projection $S_{z}$ and a combinatorial variable $k$ which
replaces the total orbital momentum $L$ and its projection $L_{z}$
used in the spherical potentials of atoms. Hund's first rule states
that $S$ should be maximal in the ground state which implies $S=\frac{N}{2}$
for $1\leq N\leq3$ and $S=\frac{6-N}{2}$ for $4\leq N\leq6$. The
states and the energy corrections due to the ferromagnetic exchange
are listed in Table \ref{tab:IsolatedDotHundEnergyCorrection}.

The ground states in our QD model are degenerate. In a real system,
this degeneracy will be lifted by a hierarchy of interactions, including
the aspherical Coulomb interaction and the spin-orbit interaction
as reflected in Hund's second and third rule, and finally by the coupling
to the leads. All these effects will affect the weights $w_{i\left(N\right)}$.

In our simple estimate we avoid selecting specific states from the
ground state manifold. We calculate the energy differences and weights
for all possible transitions from all elements of the ground state
manifold to the $N_{e}\pm1$ eigenstates and subsequently average
over ground state space, as listed in Table \ref{tab:EnergiesWeightsZeroTemp}.

\begin{table}
\begin{centering}
\begin{tabular}{|c|c|c|c|}
\hline 
$N=N_{e}-6$ & $\left(N,S,\xi\right)$ & degeneracy & $\Delta E_{Hund}$\tabularnewline
\hline 
\hline 
$1$ & ${\bf \left(1,\frac{1}{2},1\right)}$ & ${\bf 6}$ & ${\bf 0}$\tabularnewline
\hline 
\hline 
\multirow{3}{*}{$2$} & ${\bf \left(2,1,2\right)}$ & ${\bf 9}$ & ${\bf -\frac{1}{2}J}$\tabularnewline
\cline{2-4}
 & $\left(2,0,2\right)$ & $3$ & $+\frac{3}{2}J$\tabularnewline
\cline{2-4}
 & $\left(2,0,0\right)$ & $3$ & $0$\tabularnewline
\hline 
\hline 
\multirow{3}{*}{$3$} & ${\bf \left(3,\frac{3}{2},3\right)}$ & ${\bf 4}$ & ${\bf -\frac{3}{2}J}$\tabularnewline
\cline{2-4}
 & $\left(3,\frac{1}{2},3\right)$ & $4$ & $+\frac{3}{2}J$\tabularnewline
\cline{2-4}
 & $\left(3,\frac{1}{2},1\right)$ & $12$ & $0$\tabularnewline
\hline 
\hline 
\multirow{3}{*}{$4$} & ${\bf \left(4,1,2\right)}$ & ${\bf 9}$ & ${\bf -\frac{1}{2}J}$\tabularnewline
\cline{2-4}
 & $\left(4,0,2\right)$ & $3$ & $+\frac{3}{2}J$\tabularnewline
\cline{2-4}
 & $\left(4,0,0\right)$ & $3$ & $0$\tabularnewline
\hline 
\hline 
$5$ & ${\bf \left(5,\frac{1}{2},1\right)}$ & ${\bf 6}$ & ${\bf 0}$\tabularnewline
\hline 
\end{tabular}
\par\end{centering}
\caption{Contribution of Hund exchange to many-electron eigenstates of the
partially filled third shell of an isolated QD. The energies depend
on the occupation number $N=N_{e}-6$, the total spin $S$, and the
number of unpaired electrons (seniority) $\xi$. The ground states
are written in bold-face.}\label{tab:IsolatedDotHundEnergyCorrection}
\end{table}

\begin{table}
\begin{centering}
\begin{tabular}{|c|c|c|c|c|c|}
\hline 
$N$ & ground state & final state & $\epsilon_{i\left(N\right)}$  & $w_{i\left(N\right)}$  & addition energy\tabularnewline
\hline 
\hline 
\multirow{4}{*}{$1$} & \multirow{4}{*}{$\left(1,\frac{1}{2},1\right)$} & $\left(0,0,0\right)$ & $-\frac{U}{2}$ & $\frac{1}{6}$ & \multirow{4}{*}{$U$}\tabularnewline
\cline{3-5}
 &  & $\left(2,0,0\right)$ & $\frac{U}{2}$ & $\frac{1}{6}$ & \tabularnewline
\cline{3-5}
 &  & $\left(2,0,2\right)$ & $\frac{U}{2}+\frac{3}{2}J$ & $\frac{1}{6}$ & \tabularnewline
\cline{3-5}
 &  & $\left(2,1,2\right)$ & $\frac{U}{2}-\frac{1}{2}J$ & $\frac{1}{2}$ & \tabularnewline
\hline 
\hline 
\multirow{4}{*}{2} & \multirow{4}{*}{$\left(2,1,2\right)$} & $\left(1,\frac{1}{2},1\right)$ & -$\left(\frac{U}{2}+\frac{1}{2}J\right)$ & $\frac{1}{3}$ & \multirow{4}{*}{$U+\frac{2}{3}J$}\tabularnewline
\cline{3-5}
 &  & $\left(3,\frac{1}{2},1\right)$ & $\frac{U}{2}+\frac{1}{2}J$ & $\frac{1}{3}$ & \tabularnewline
\cline{3-5}
 &  & $\left(3,\frac{1}{2},3\right)$ & $\frac{U}{2}+2J$ & $\frac{1}{9}$ & \tabularnewline
\cline{3-5}
 &  & $\left(3,\frac{3}{2},3\right)$ & $\frac{U}{2}-J$ & $\frac{2}{9}$ & \tabularnewline
\hline 
\hline 
\multirow{2}{*}{$3$} & \multirow{2}{*}{$\left(3,\frac{3}{2},3\right)$} & $\left(2,1,2\right)$ & $-\left(\frac{U}{2}+J\right)$ & $\frac{1}{2}$ & \multirow{2}{*}{$U+2J$}\tabularnewline
\cline{3-5}
 &  & $\left(4,1,2\right)$ & $\frac{U}{2}+J$ & $\frac{1}{2}$ & \tabularnewline
\hline 
\end{tabular}
\par\end{centering}
\caption{Energies $\epsilon_{i\left(N\right)}$ and weights $w_{i\left(N\right)}$
appearing in the spectral function of an isolated QD, resulting averaged
energy $\bar{E}\left(N\right)$ of the charge peaks and addition energy}\label{tab:EnergiesWeightsZeroTemp}
\end{table}

As the temperature is raised the excited Hund multiplets become thermally
populated which leads to changes in the addition energies. The energies
and probabilities are listed in Table \ref{tab:EnergiesWeightsHalfFillingFiniteTemp}.
For $k_{B}T\simeq J\ll U$, we calculate 
\begin{equation}
\bar{E}_{c}\left(T\right)=2\sum_{i}\left|\epsilon_{i}\right|w_{i}p_{i}\left(T\right)=U+2J\frac{4\left(1-e^{-3J/(k_{B}I)}\right)}{Z_{H}\left(T\right)}
\end{equation}
 with the Hund's rule contribution to the partition function 
\begin{equation}
Z_{H}\left(T\right)=4+12e^{-\frac{3}{2}\frac{J}{k_{B}T}}+4e^{-3\frac{J}{k_{B}T}}\quad.
\end{equation}

\begin{table}[h]
\begin{centering}
\begin{tabular}{|c|c|c|c|c|}
\hline 
initial state & final state & $\epsilon_{i}$ & $w_{i}$ & $p_{i}\left(T\right)$\tabularnewline
\hline 
\hline 
\multirow{2}{*}{$\left(3,\frac{3}{2},3\right)$} & $\left(2,1,2\right)$ & $-\left(\frac{U}{2}+J\right)$ & $2$ & \multirow{2}{*}{$\frac{1}{Z_{H}\left(T\right)}$}\tabularnewline
\cline{2-4}
 & $\left(4,1,2\right)$ & $\frac{U}{2}+J$ & $2$ & \tabularnewline
\hline 
\hline 
\multirow{6}{*}{$\left(3,\frac{1}{2},1\right)$} & $\left(2,1,2\right)$ & $-\left(\frac{U}{2}-\frac{1}{2}J\right)$ & $3$ & \multirow{6}{*}{$\frac{e^{\left(-\frac{3}{2}\frac{J}{k_{B}T}\right)}}{Z_{H}\left(T\right)}$}\tabularnewline
\cline{2-4}
 & $\left(2,0,2\right)$ & $-\left(\frac{U}{2}+\frac{3}{2}J\right)$ & $1$ & \tabularnewline
\cline{2-4}
 & $\left(2,0,0\right)$ & $-\frac{U}{2}$ & $2$ & \tabularnewline
\cline{2-4}
 & $\left(4,1,2\right)$ & $\left(\frac{U}{2}-\frac{1}{2}J\right)$ & $3$ & \tabularnewline
\cline{2-4}
 & $\left(4,0,2\right)$ & $\left(\frac{U}{2}+\frac{3}{2}J\right)$ & $1$ & \tabularnewline
\cline{2-4}
 & $\left(4,0,0\right)$ & $\frac{U}{2}$ & $2$ & \tabularnewline
\hline 
\hline 
\multirow{2}{*}{$\left(3,\frac{3}{2},3\right)$} & $\left(2,1,2\right)$ & $-\left(\frac{U}{2}-J\right)$ & $2$ & \multirow{2}{*}{$\frac{e^{\left(-3\frac{J}{k_{B}T}\right)}}{Z_{H}\left(T\right)}$}\tabularnewline
\cline{2-4}
 & $\left(4,1,2\right)$ & $\left(\frac{U}{2}+J\right)$ & $2$ & \tabularnewline
\hline 
\end{tabular}
\par\end{centering}
\caption{Energies $\epsilon_{i}$, weights $w_{i}$, and Boltzmann factors
appearing in the finite-T averaged energy $\bar{E}_{c}\left(N\right)$
addition energy at half-filling. }\label{tab:EnergiesWeightsHalfFillingFiniteTemp}
\end{table}

\subsection{Rotationally Invariant Slave Boson (RISB) approach}

The central focus is the low-energy many-body (Kondo) resonance in
the low-temperature spectral function for the model Hamiltonian 
\begin{equation}
H=H_{dot}+H_{hyb}+H_{leads}
\end{equation}
where
\begin{eqnarray}
H_{dot} &= &\sum_{\nu\sigma}\left[\epsilon_{d}n_{d\nu\sigma}+\frac{1}{2}Un_{d\nu\sigma}n_{d\nu-\sigma}\right]\nonumber\\
 & &
+\frac{1}{2}\text{\ensuremath{\sum}}_{\nu\neq\nu',\sigma\sigma'}\left[U'n_{d\nu\sigma}n_{d\nu'\sigma'}-2J{\bf S}_{d\nu}\cdot{\bf S}_{d\nu'}\right]\nonumber\\
H_{hyb} & = &\sum_{k \nu \sigma}\mathcal{\mathscr{V}}_{\nu}\left(c_{k\nu\sigma}^{\dagger}d_{\nu\sigma}+d_{\nu\sigma}^{\dagger}c_{k\nu\sigma}\right)\nonumber\\
H_{leads} & = & \sum_{k \nu \sigma}\epsilon_{k}c_{k\nu\sigma}^{\dagger}c_{k\nu\sigma} \quad .
\end{eqnarray}
We use the version of the Rotationally Invariant Slave Boson (RISB) approach
to treat the Hund exchange interaction. Our notation closely follows Ref.
\cite{Lechermann2007}.

We begin by expressing the annihilation operators for dot electrons,
\begin{align}
d_{\nu\sigma} & =R_{\nu\sigma}^{\dagger}\left[\Phi\right]f_{\nu\sigma}\label{eq:SlaveBosonTrafo}
\end{align}
by combinations of bosonic operators $\Phi=\left\{ \ldots,\phi_{\Gamma n},\ldots\right\} $
and pseudo-fermion operators $f_{\nu\sigma}^{\dagger}$. The latter
refer to the quasiparticle degrees of freedom of the local Fermi liquid. The indices  of the boson operators $\phi _{\Gamma n}$, 
 $n$ and $\Gamma$, refer to the usual Fock states and the 
multiplet states $\left|\Gamma\right\rangle =\left|d^{N_{e}};SS_{z}\right\rangle $ and 
of the isolated QD , respectively. Here $N_{e}=N-6$ is the occupancy
of the partially filled third shell, $S$ is the total spin and $S_{z}$
is the projection along the quantization axis. 

The operator $R_{\nu\sigma}^{\dagger}\left[\Phi\right]$
is given by
\begin{align}
R_{\nu\sigma}^{\dagger}\left[\Phi\right] & =\frac{\hat{\gamma}_{\nu\sigma}\left[\Phi\right]}{\sqrt{\hat{n}_{\nu\sigma}\left[\Phi\right]\left(1-\hat{n}_{\nu\sigma}\left[\Phi\right]\right)}}\nonumber \\
\hat{\gamma}_{\nu\sigma}\left[\Phi\right] & =\sum_{\Gamma\Gamma',nn'}\left\langle \Gamma\left|d_{\nu\sigma}^{\dagger}\right|\Gamma'\right\rangle \left\langle n\left|f_{\nu\sigma}^{\dagger}\right|n'\right\rangle \hat{n}_{\nu\sigma}\left[\Phi\right]\nonumber \\
\hat{n}_{\nu\sigma}\left[\Phi\right] & =\sum_{\Gamma n}\left\langle n\left|f_{\nu\sigma}^{\dagger}f_{\nu\sigma}\right|n\right\rangle \phi_{\Gamma n}^{\dagger}\phi_{\Gamma n} \quad  . \label{eq:SlaveBosonNormalization}
\end{align}

\begin{figure}[h]
\begin{centering}
\includegraphics[width=0.6\columnwidth]{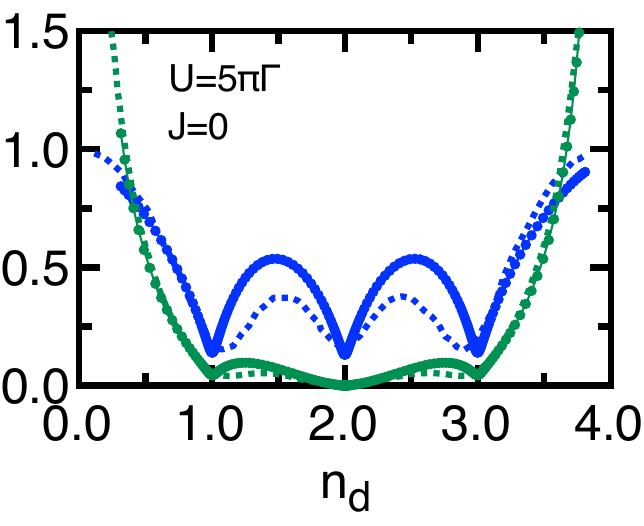}
\caption{Variation with dot occupation ($n_d$) of the renormalized parameters $\left|\tilde{\epsilon}_{d}\right|$ (green) and $\tilde{\Gamma}/ \Gamma$ (blue) for the Constant Interaction model $J=0$ as deduced from NRG (dotted lines) and RISB (filled dots). Following \cite{Kleeorin2018}, the bare hybridization strengths are replaced according to $\Gamma=\Gamma_{NRG}\to2\Gamma_{RISB}$.}
 \label{fig:ComparisonTwoOrbitalCIModel} 
\par\end{centering}
\end{figure}

To eliminate non-physical states from the Hilbert space we introduce
the constraints
\begin{align}
\sum_{\Gamma n}\phi_{\Gamma n}^{\dagger}\phi_{\Gamma n} & =1\nonumber \\
\sum_{\Gamma n}\left\langle n\left|f_{\nu\sigma}^{\dagger}f_{\nu\sigma}\right|n\right\rangle \phi_{\Gamma n}^{\dagger}\phi_{\Gamma n}= & f_{\nu\sigma}^{\dagger}f_{\nu\sigma}
\label{eq:RISBConstraints}
\end{align}
which ensure completeness and conservation of quasiparticle numbers. 

We adopt the saddle-point approximation replacing the boson fields
by their expectation values
\begin{equation}
\phi_{\Gamma n} \to \varphi_{\Gamma n} \quad .
\end{equation}
The latter have to be determined self-consistently
by minimizing the free-energy 
\begin{equation}
F_{MF}=-\frac{1}{\beta}\ln\mbox{Tr}\exp\left\{ -\beta H_{MF}\right\} 
\end{equation}
where the constraints Eq. (\ref{eq:RISBConstraints}) are included by Lagrange parameters $\lambda$ and $\Lambda_{\nu \sigma}$. This yields
\begin{align}
\lefteqn{H_{MF} = } \nonumber\\
& \sum_{k \nu \sigma}\epsilon_{k}c_{k\nu\sigma}^{\dagger}c_{k\nu\sigma} +\sum_{k \nu \sigma} \left\{ R_{\nu\sigma}^{\dagger} \left[\Phi\right]\mathcal{V}_{\nu} c_{k\nu\sigma}^{\dagger}f_{\nu\sigma}+h.c.\right\} \nonumber\\
 & -\sum_{\nu\sigma}\Lambda_{\nu\sigma}f_{\nu\sigma}^{\dagger}f_{\nu\sigma} \nonumber\\
 & + \sum_{\Gamma n}
\left\{ E_{\Gamma} +\sum_{\nu \sigma}\Lambda_{\nu \sigma} \left\langle n\left|f_{\nu \sigma}^{\dagger}f_{\nu \sigma}\right| n\right\rangle \right\}  
\left| \varphi_{\Gamma n}\right|^2 
\nonumber \\  
 & + \lambda\left\{ \sum_{\Gamma n}\left| \varphi_{\Gamma n}\right|^2-1\right\}  \quad.
 \label{eq:MFHamiltonian}
\end{align}

The calculation of the free-energy is rather straightforward since
the fermionic part of $H_{MF}$ corresponds to a non-interacting multi-orbital Anderson
impurity, i. e., a set of  Resonant Level models with renormalized centers of gravity
$-\Lambda_{\nu\sigma}$ and hybridization $\tilde{\mathcal{V}}_{\nu\sigma}=R_{\nu\sigma}^{\dagger}\left[\Phi\right]\mathcal{V}_{\nu}$.
The evaluation of the dot contribution to the ground state energy
proceeds in close analogy to the derivation \citet{HewsonBook}

We neglect a potential orbital dependence of the hybridization
$\mathcal{V}_{\nu}\to\mathcal{V}$ and in the Hamiltonian and assume
an isotropic quasiparticle renormalization with a matrix 
\begin{equation}
{\bf z}\left[\Phi\right]={\bf R}^{\dagger}\left[\Phi\right]{\bf R}\left[\Phi\right]=z\left[\Phi\right]\mathbb{I}
\end{equation}
proportional to the identity. In addition, we neglect Crystal
Field terms requiring the same center of gravity for all $\nu$-channels,
i. e., $-\Lambda_{\nu\sigma}\to\tilde{\epsilon}_{d}$. 
To further reduce the number of variational parameters, we approximate
the boson expectation values by their form at the saddle point 
\begin{equation}
\varphi_{\Gamma n}\to\left\langle \Gamma\vert n\right\rangle \mathfrak{\mathcal{Y}}_{\Gamma}\quad.
\end{equation}

\begin{figure}[h]
\includegraphics[width=0.6\columnwidth]{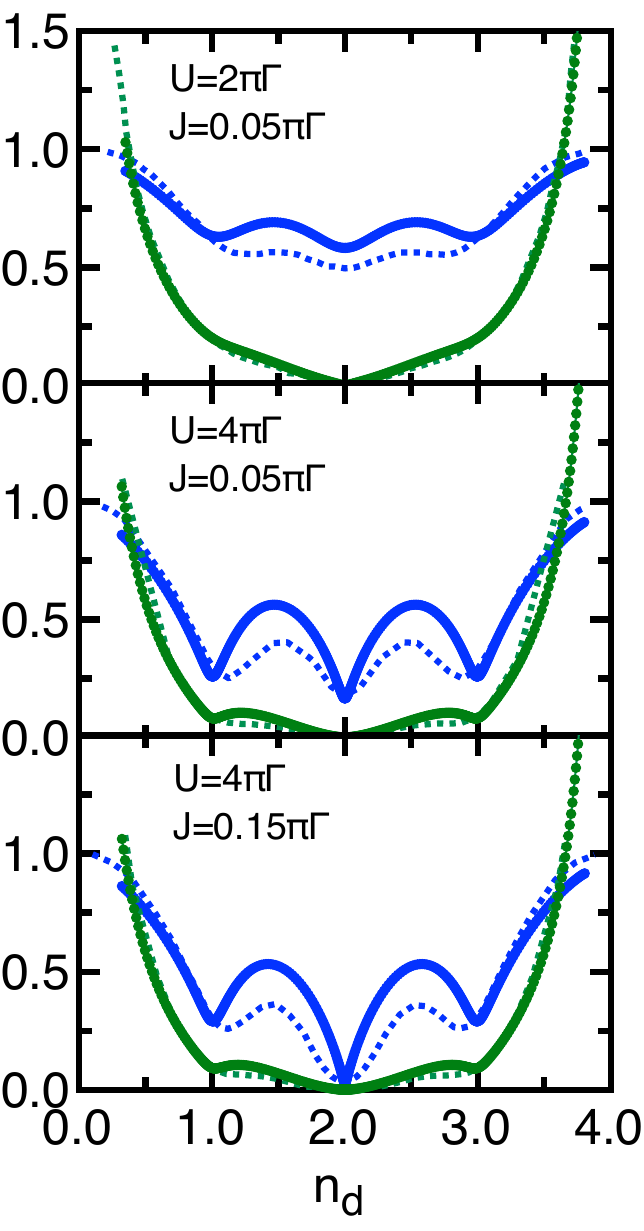}
\caption{Variation with dot occupation ($n_d$) of the renormalized parameters $\left|\tilde{\epsilon}_{d}\right|$ (green) and $\tilde{\Gamma}/ \Gamma$ (blue) as deduced from NRG (dotted lines) and RISB (filled dots), calculated with the Kanamori parametrization for the inter-orbital Coulomb repulsion. Following , the bare hybridization strengths are replaced according to $\Gamma=\Gamma_{NRG}\to2\Gamma_{RISB}$.}
\label{fig:ComparisonTwoOrbitalKanamoriParametrization}
\end{figure}

To assess the validity of the RISB approximation, we compare its predictions
for the centers $\tilde{\epsilon}_{d}$ and and width renormalization
$\tilde{\Gamma}/ \Gamma$ of the Kondo resonances of a two-channel
Anderson model to their counterparts derived from the numerically
exact NRG by \citet{Nishikawa2010b}. NRG reference data are available
for CI models ($J=0$) and models with finite Hund exchange $J>0$ and Kanamori 
parametrization $U'=U-\frac{3}{2}J$ for the inter-orbital Coulomb repulsion.
For this reason, we also employ the Kanamori parametrization for
the inter-orbital Coulomb repulsion $U'$ in our RISB calculation.
The comparison for both the CI model in the strong-coupling case in
Figure \ref{fig:ComparisonTwoOrbitalCIModel} and the Kanamori model
for fixed ratio $J/U$ but different correlation strengths in Figure
\ref{fig:ComparisonTwoOrbitalKanamoriParametrization} clearly show
that the NRG results are reasonably well reproduced by RISB provided
the hybridization strengths are replaced $\Gamma=\Gamma_{NRG}\to 2\Gamma_{RISB}$ as
suggested by \cite{Kleeorin2018} which implies $\left(\frac{U}{\Gamma}\right)_{RISB}=2\left(\frac{U}{\Gamma}\right)_{NRG}$.

\end{document}